\documentclass[twocolumn,aps,prl,showpacs,groupedaddress]{revtex4}

\usepackage{amsmath}
\usepackage{dcolumn}
\usepackage{epsfig}
\usepackage{graphicx}
\usepackage{latexsym}

\usepackage{epstopdf}
\begin{document}

\title{Speckle Imaging of Spin Fluctuations in a Strongly Interacting Fermi Gas}

\author{Christian Sanner}
\author{Edward J. Su}
\author{Aviv Keshet}
\author{Wujie Huang}
\author{Jonathon Gillen}
\author{Ralf Gommers}
\author{Wolfgang Ketterle}
\affiliation{MIT-Harvard Center for Ultracold Atoms, Research
Laboratory of Electronics, and Department of Physics, Massachusetts
Institute of Technology, Cambridge MA 02139}

\begin{abstract}
Spin fluctuations and density fluctuations are studied for a
two-component gas of strongly interacting fermions along the BEC-BCS
crossover. This is done by in-situ imaging of dispersive speckle
patterns. Compressibility and magnetic susceptibility are determined
from the measured fluctuations. This new sensitive method easily
resolves a tenfold suppression of spin fluctuations below shot noise
due to pairing, and can be applied to novel magnetic phases in
optical lattices.
\end{abstract}


\maketitle

One frontier in the field of ultracold atoms is the realization of
quantum systems with strong interactions and strong correlations.
Many properties of strongly correlated systems cannot be deduced
from mean density distributions. This has drawn interest toward novel ways of
probing cold atoms, e.g. via RF spectroscopy \cite{MITRF,JILARF},
Bragg and Raman scattering \cite{MITBragg},
interferometric methods \cite{DemlerInterference,ZoranInterference}
and by recording density correlations
\cite{AspectHBT,Bloch06NoiseFermion,Jin05PairCorrelation}. Further
insight into quantum systems is obtained by looking not only at
expectation values, but also at fluctuations. Several recent studies
looked at density fluctuations, either of bosons around the
superfluid-to-Mott insulator transition
\cite{Chin09Incompressibility,GreinerMottSingle,BlochMottSingle}, or
of a gas of non-interacting fermions
\cite{EsslingerFermiNoise,KetterleFermiNoise}.

In this paper, we extend the study of fluctuations of ultracold
gases in several ways. First, we apply it to a two-component Fermi
gas across the BEC-BCS crossover. Second, we implement a very
sensitive way to measure fluctuations in the magnetization, i.e. the
difference of the densities in the two different states. Third,
we introduce the technique of speckle imaging as a simple and highly
sensitive method to characterize fluctuations.

Our work is motivated by the prospect of realizing wide classes of
spin Hamiltonians using a two-component gas of ultracold atoms in an
optical lattice \cite{DuanSpinExchange,KuklovLattice}. An important
thermodynamic quantity to characterize two-component systems is the
spin susceptibility, which provides a clear signature of phase
transitions or crossovers involving the onset of pairing or magnetic
order \cite{StringariSpinFluc,NoiseSpectroscopyDemler}. At a
ferromagnetic phase transition the susceptibility diverges, whereas
in a transition to a paired or antiferromagnetic phase the
susceptibility becomes exponentially small in the ratio of the pair
binding energy (or antiferromagnetic gap) to the temperature. The
fluctuation-dissipation theorem relates response functions to
fluctuations, consequently the spin susceptibility can be determined
by measuring the fluctuations in the relative density of the two
spin components.

\begin{figure}[tbh!]
\begin{center}
\includegraphics[width=1.0\columnwidth]{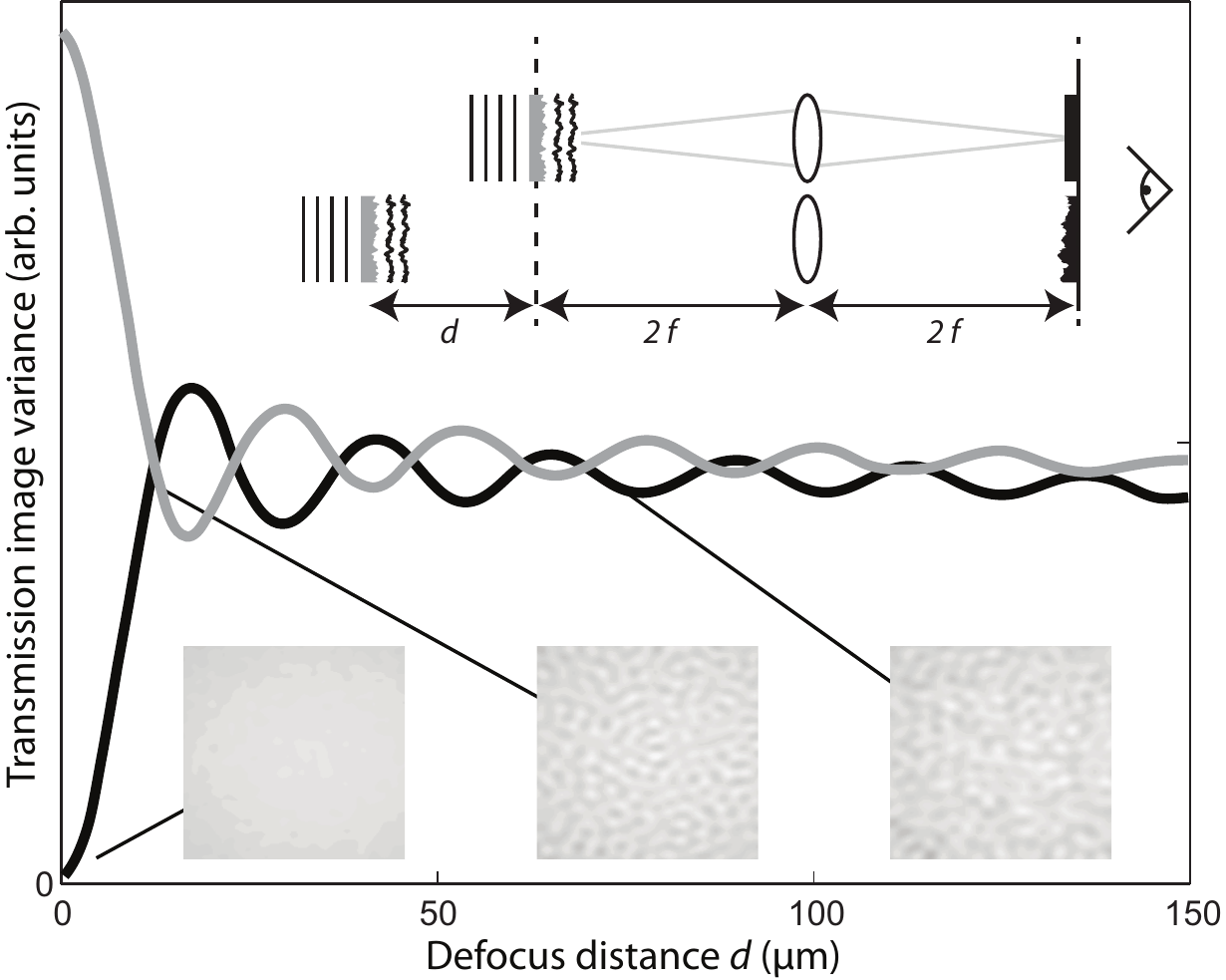}
\caption[]{Simulation of propagation effects after light has passed through
a Poissonian phase noise object. Shown are the variance measured in
the amplitude or in-phase quadrature (black line) and the
out-of-phase quadrature (gray line) as a function of defocus
distance, for an imaging system with a numerical aperture of 0.14.
Within a distance less than $5$ percent of our cloud size, noise
becomes equally distributed between the two quadratures and the
variances in transmission and phase-contrast images become the same.
(Top inset) For small phase fluctuations, an in-focus phase noise
object gives no amplitude contrast, but when it is out of focus it
does. (Bottom inset) Sample intensity patterns for a defocused phase
object. \label{f:specklesim}}
\end{center}
\end{figure}

In our experiment spin fluctuations create phase shifts of (detuned) imaging light that vary randomly in space;
we measure them by imaging the resulting speckle patterns. When imaging atom clouds,
one usually relates the transmitted light intensity with absorption
and its phase with dispersion \cite{Ketterle98Varenna}. This is
different in our method. Spin and density fluctuations occur on all spatial
scales down to the interatomic separation, and their observation is
limited by the maximum resolution of the imaging system. For typical
atom clouds in 3D (in contrast to 2D experiments
\cite{Chin09Incompressibility,GreinerMottSingle,BlochMottSingle} or
very small clouds in 3D \cite{EsslingerFermiNoise}) the cloud size
is larger than the Rayleigh range corresponding to the maximum
resolution of the imaging systems. Therefore, for the smallest
resolvable fluctuations, the depth of field is smaller than
the sample size and, necessarily, the recorded image is modified by
propagation effects. Propagation mixes up amplitude and phase. This
can be easily seen in the case of a phase grating, which creates an
interference pattern further downstream. Conversely, due to Fermat's principle, 
for an amplitude object rephasing takes place only in conjugate
planes, and therefore out of focus there is a phase-contrast signal.  
Similar physics is responsible for laser speckle when a rough surface scatters light with random phases
\cite{Goodman07}, and occurs when a Bose-Einstein condensate with
phase fluctuations develops density fluctuations during ballistic
expansion \cite{ErtmerPRL}, or when a phase-contrast signal is
turned into an amplitude signal by deliberate defocusing
\cite{ScholtenDefocusContrast}.

In our experiments, we use off-resonant imaging light due to the
high optical density of the cloud. Absorption decreases inversely
proportional to the detuning squared, whereas the dispersive signal
falls off more slowly, inversely proportional to the detuning.
Therefore, the dominant signal is initially dispersive, but is
converted to an amplitude signal during propagation.  Simulations
using Poissonian noise confirm this picture
(Fig.~\ref{f:specklesim}): After a short propagation distance, the
power of phase and amplitude noise are equal, independently of
whether the noise was purely dispersive or absorptive before
propagation. This feature of speckle makes our imaging technique
both simple and robust. It is insensitive against defocusing, and
allows us to image fluctuations of the real part of the refractive
index (i.e. a phase signal) without a phase plate or other Fourier
optics.

The experiments were performed with typically $10^6$ $^6$Li atoms in
each of the two lowest hyperfine states $|1\rangle$ and $|2\rangle$
confined in an optical dipole trap oriented at $45^{\circ}$ to the
imaging axis with radial and axial trap frequencies
$\omega_{r}=2\pi\times108.9(6)$~s$^{-1}$ and
$\omega_{z}=2\pi\times7.75(3)$~s$^{-1}$. For the samples imaged at
527G, the sample preparation was similar to that described in
\cite{KetterleFermiNoise}, with a temperature of $0.14(1) T_F$. The
samples imaged at other magnetic fields were prepared in a similar
fashion, except that evaporation was performed at 1000G to a final
temperature of $T=0.13(1) T_F$ before ramping the magnetic field
over $1.5 s$ to its final value. The temperature at 1000G was
determined by fitting a noninteracting Thomas-Fermi distribution in
time of flight. The temperatures at other points in the crossover
were related to that value assuming an isentropic ramp, using
calculations presented in \cite{LevinCondensateFraction}. Using this
method we obtain temperatures of $0.13(1) T_F$ at 915G, $0.19(1)
T_F$ at 830G, and $0.19(3) T_F$ at 790G where additional evaporation
was performed to achieve a central optical density similar to that
at the other magnetic fields. The extent of the cloud along the
imaging direction was 135$\mu$m, much larger than the Rayleigh range
of 8$\mu$m for our imaging system with a numerical aperture of 0.14.

\begin{figure}[tbh!]
\begin{center}
\includegraphics[width=1.0\columnwidth]{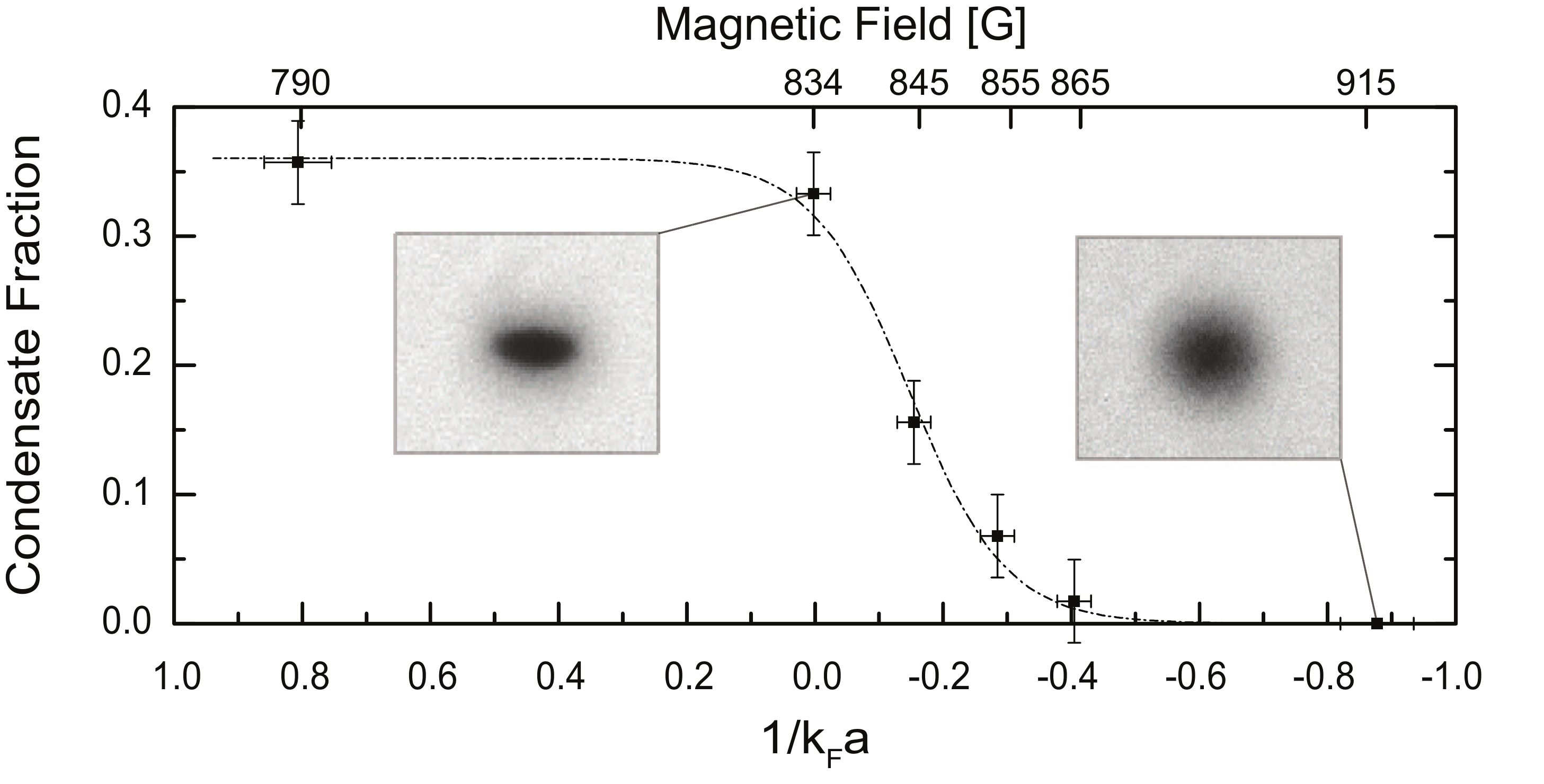}
\caption[]{Measured condensate fraction as a function of
dimensionless interaction strength $1/(k_{F}a)$. Insets show typical
images from which the condensate fraction was extracted by fitting a
bimodal distribution. The dashed line is a sigmoidal fit to guide
the eye. \label{f:condensatefraction}}
\end{center}
\end{figure}

The superfluid to normal phase boundary was determined by measuring
condensate fraction (Fig.~\ref{f:condensatefraction}) using the
standard magnetic field sweep technique \cite{JILASweep,MITSweep}.
For this, the magnetic field was rapidly switched to 570G to
transfer atom pairs to more deeply bound pairs (molecules) which
survive ballistic expansion. For resonant imaging of the molecules,
the field was ramped back to 790G over 10 ms. The condensate
fraction was determined by fitting the one-dimensional density
profiles with a bimodal distribution.

\begin{figure}[tbh!]
\begin{center}
\includegraphics[width=1.0\columnwidth]{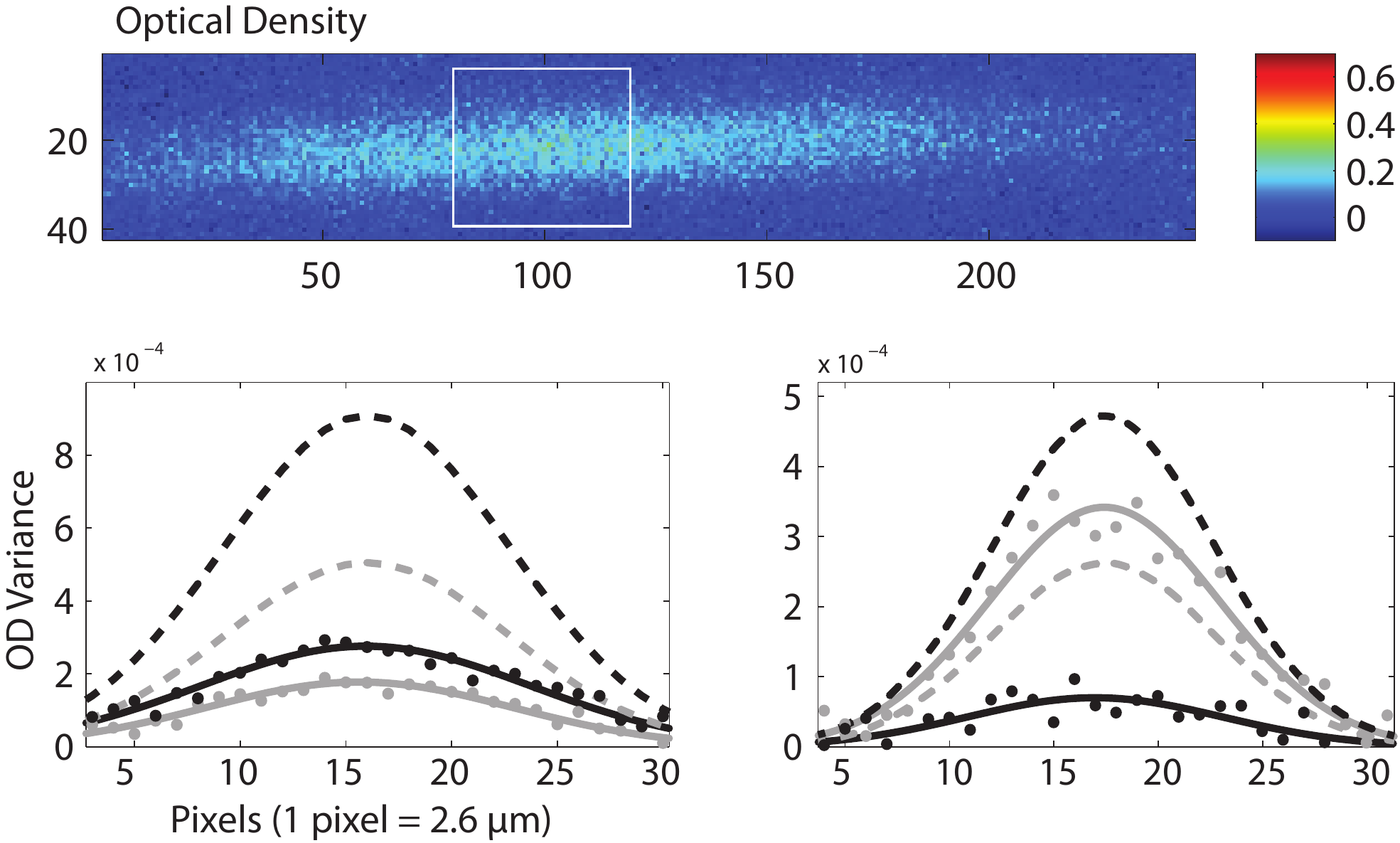}
\caption[]{(Color online) (Top) Example speckle noise image, with white box
indicating analysis region. (Bottom) Noise data for noninteracting (left) and resonantly interacting (right) cold clouds, 
showing $\Delta_-^2$ (black dots) and $\Delta_+^2$ (grey dots). Solid lines are Gaussian fits to the data,
and dotted lines are expected full Poissonian noise for the
corresponding quantities based on density determined from
off-resonant absorption. \label{f:rawnoise}}
\end{center}
\end{figure}

As previously described, propagation converts spatial fluctuations
in the refractive index into amplitude fluctuations on the detector.
For different choices of the probe light frequency, the two atomic
spin states will have different real polarizabilities and the local
refractive index will be a different linear combination of the
(line-of-sight integrated) column densities $n_1$ and $n_2$. To
measure the susceptibility we choose a probe light frequency exactly
between the resonances for states $|1\rangle$ and $|2\rangle$, so
that the real polarizabilities are opposite and the refractive index
is proportional to the magnetization $(n_1 - n_2)$. The intensity
fluctuations on the detector after propagation are consequently
proportional to the fluctuations in magnetization. Since a
refractive index proportional to $(n_1 + n_2)$ occurs only in the
limit of infinite detuning, we measure the fluctuations in the total
density by exploiting the fact that the fluctuations in total
density can be inferred from the fluctuations in two different
linear combinations of $n_1$ and $n_2$. For convenience, we obtain
the second linear combination using a detuning that has the same
value, but opposite sign for state $|2\rangle$, and therefore three
times the value for state $|1\rangle$. With this detuning, we record
images of the fluctuations in $(n_1/3 + n_2)$.

In principle this information can be obtained by taking separate
absorption images on resonance for states $|1\rangle$ and
$|2\rangle$. However, the images would have to be taken on a
timescale much faster than that of atomic motion. and for there would
 be increased technical noise from the subtraction of large numbers. 
The use of dispersive imaging has the additional advantage over absorption in that the number of
scattered photons in the forward direction is enhanced by
superradiance. As a result, for the same amount of heating, a larger
number of signal photons can be collected \cite{Ketterle98Varenna}.
This is crucial for measuring atomic noise, which requires the
collection of several signal photons per atom. The choice of
detuning between the transitions of the two states has the important
feature that the index of refraction for an equal mixture fluctuates
around zero, avoiding any lensing and other distortions of the probe
beam. This is not the case for other choices of detuning, and
indeed, we observe some excess noise in those images (see below). At
the detunings chosen, 10 percent residual attenuation
is observed, some due to off-resonant absorption, some due to
dispersive scattering of light out of the imaging system by small
scale density fluctuations. The contribution to the variance  of the absorption
signal relative to the dispersive signal scales as $(2\Gamma)^2/\delta^2\approx0.006$ and can be neglected in
the interpretation of the data.

The noise analysis procedure was nearly identical to that performed
in \cite{KetterleFermiNoise}. A high-pass filter with a cutoff wavelength of 13
$\mu m$ was applied to each image of the cloud to minimize the effect of
fluctuations in total atom number. Then, for each pixel position,
the variance of the optical densities at that position in the
different images was computed. After the subtraction of the
contribution of photon shot noise, the resulting variance image
reflects the noise contribution from the atoms. 

\begin{figure}[tbh!]
\begin{center}
\includegraphics[width=1.0\columnwidth]{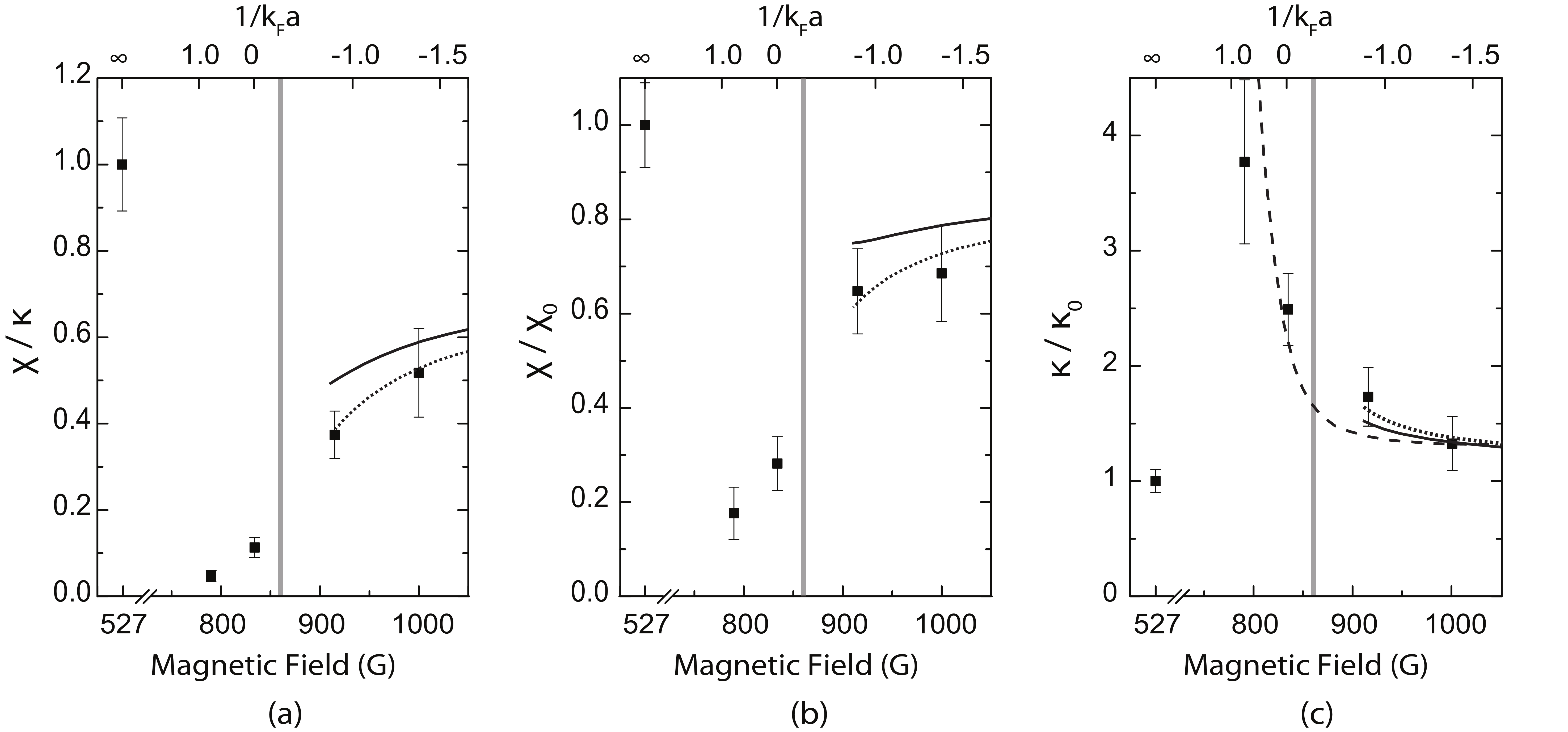}
\caption[]{(a) The ratio $\chi/\kappa$, (b) the normalized
susceptibility $\chi/\chi_0$, and (c) the normalized compressibility
$\kappa/\kappa_0$ in the BEC-BCS crossover. The variances derived
from sequences of images are converted into thermodynamic variables
using the measured temperatures and a calibration factor determined
from the noninteracting gas. The vertical line indicates the onset
region of superfluidity, as determined via condensate fraction
measurements. The curves show theoretical zero temperature estimates
based on 1st (dotted) and 2nd order (solid) perturbative formulas
obtained from Landau's Fermi-liquid theory integrated along the line
of sight, and results from a Monte Carlo calculation (dashed) for
the compressibility in a homogeneous system \cite{AstrCompr}.
\label{f:chikappa}}
\end{center}
\end{figure}

The goal of our noise measurements is to determine at various
interaction strengths the normalized susceptibility
$\widetilde{\chi}=\chi/\chi_0$ and compressibility
$\widetilde{\kappa}=\kappa/\kappa_0$, where $\chi_0=3n/2E_F$ and
$\kappa_0=3/2nE_F$ are the susceptibility and compressibility of a
zero-temperature non interacting Fermi gas of the same total density
$n$ and Fermi energy $E_F$. Before studying spin fluctuations
through the BEC-BCS crossover, we therefore calibrate our
measurement by measuring the spin fluctuations in a noninteracting
mixture, realized at 527G where the scattering length between the
two states vanishes. Fig.~\ref{f:rawnoise} shows raw profiles of the
difference and sum variances of the measured optical densities $\Delta_-^2=(c\,\Delta(N_1-N_2))^2$ and
$\Delta_+^2=(c'\,\Delta(N_1/3+N_2))^2$. In these relations $c$ and
$c'$ are the conversion factors between number fluctuations
and fluctuations in optical density in the specific
probe volume $V$. Without interactions, $N_1$ and $N_2$ are
uncorrelated, and one predicts
$(\Delta(N_1-N_2))^2/(\Delta(N_1/3+N_2))^2=2/(1+(1/3)^2)=1.8$. The
observed ratio of $\Delta_-^2/\Delta_+^2=1.56(14)$ reflects 
excess noise contributing to $\Delta_+^2$ due to residual systematic
dispersive effects and is accounted for by setting
$c'/c=\sqrt{1.8/1.56}$. For high temperatures, the atomic noise of
the non-interacting gas approaches shot noise; for lower temperatures
we observe a reduction in noise due to Pauli blocking as in our
previous work \cite{KetterleFermiNoise}. With our new method, we
easily discern spin fluctuations with a variance of less than 10
percent of atom shot noise.

The fluctuation dissipation theorem connects the variances
$(\Delta(N_1-N_2))^2$ and $(\Delta(N_1+N_2))^2$ to the
susceptibility $\widetilde{\chi}$ and the compressibility
$\widetilde{\kappa}$ via $(\Delta(N_1-N_2))^2=3N/2 \; (T/T_{F}) \;
\widetilde{\chi}$ and $(\Delta(N_1+N_2))^2=3N/2 \; (T/T_{F}) \;
\widetilde{\kappa}$ with $N=N_1+N_2$ and $T/T_F$ being the
temperature measured in units of the Fermi temperature $T_F$.
Recomposing the variances from the two experimentally accessible
linear combinations these relations become $\Delta_-^2/Nc^2=3/2 \;
(T/T_{F}) \; \widetilde{\chi}$ and $9/4 \: \Delta_+^2/Nc'^2-1/4 \:
\Delta_-^2/Nc^2=3/2 \; (T/T_{F}) \; \widetilde{\kappa}$ and allow us
to determine $c$ and $c'$ by using the 527G noise measurements for
which $\widetilde{\chi}=\widetilde{\kappa}=1+O\left(\left(T/T_F\right)^2\right)$. This 
analysis ignores corrections due to line-of-sight integration. 


Fig.~\ref{f:chikappa} shows the spin susceptibility, the
compressibility, and the ratio between the two quantities for the
interacting mixtures as the interaction strength is varied through
the BEC-BCS crossover. The susceptibility and compressibility
reproduce the expected qualitative behavior: for the sample at
unitarity, where the transition temperature is sufficiently high
that a sizable portion of the sample is superfluid, and for the
sample on the BEC side, where even the normal-state atoms form
molecules, the spin susceptibility is strongly suppressed relative
to the compressibility. This reflects the fact that the atoms form
bound molecules or generalized Cooper pairs; the spin susceptibility
should be exponentially small in the binding energy, while the
enhanced compressibility reflects the bosonic character of the
molecular condensate. At 915G and 1000G, where the sample is above
the superfluid critical temperature, the susceptibility is larger
but still below its value for the noninteracting gas, reflecting the
persistence of pair correlations even in the normal phase of the
gas.

Above the Feshbach resonance, for attractive interactions, we
compare our results to first and second order perturbation theory in
the small parameter $k_{F}a$. This ignores the instability to the
superfluid BCS state at exponentially small temperatures. The
perturbation theory is often formulated for the Landau parameters
for a Fermi liquid \cite{StringariSpinFluc,LandauStatMech}. The
susceptibility and compressibility are given by
$\chi_{0}/\chi=(1+F^{a}_0)m/m^\ast$,
$\kappa_{0}/\kappa=(1+F_0^{s})m/m^\ast$, where $m^\ast=m(1+F_1^{s}/3)$ is
the effective mass, and $F_{l}^{s}$, $F_{l}^{a}$ are the $l$-th
angular momentum symmetric and antisymmetric Landau parameters,
respectively. Although the experimental data are taken for
relatively strong interactions outside the range of validity for a
perturbative description, the predictions still capture the trends
observed in the normal phase above the Feshbach resonance. This
shows that more accurate measurements of the susceptibility, and a
careful study of its temperature dependence, are required to reveal
the presence of a possible pseudogap phase.

In our analysis we have neglected quantum fluctuations which are
present even at zero temperature
\cite{AstrQuantumNoise,StringariSpinFluc}. They are related to the large-$q$ static structure factor 
$S(q)$ measured in \cite{HannafordStructureFactor} and proportional to
the surface of the probe volume, scaling with $N^{2/3} \log{(N)}$.
For fluctuations of the total density, their relative contribution
is roughly $N^{-1/3}/(T/T_F)$, and at most 40 percent for our
experimental parameters. Attractive interactions and pairing
suppress both the thermal and quantum spin fluctuations, but it is
not known at what temperature quantum fluctuations become essential.

Spin susceptibilities can also be obtained from the equation of
state which can be determined by analyzing the average density
profiles of imbalanced mixtures \cite{SalomonEquationOfState}. Our method has 
the advantage of being possible to implement without imbalance, and requires only local thermal equilibrium.
 Moreover fluctuations can be compared with susceptibilites determined from the equation of state to perform
absolute, model-independent thermometry for strongly interacting systems \cite{DeMarcoReview}.

In conclusion, we have demonstrated a new technique to determine
spin susceptibilities of ultracold atomic gases using speckle
imaging. We have validated and calibrated this technique using an
ideal Fermi gas and applied it to a strongly interacting Fermi gas
in the BEC-BCS crossover.  This technique is directly applicable to
studying pairing and magnetic ordering of two-component gases in
optical lattices.

We acknowledge Qijin Chen and Kathy Levin for providing calculations
of condensate fraction, Gregory Astrakharchik and Stefano Giorgini
for providing Monte Carlo results for the compressibility, Sandro
Stringari and Alessio Recati for discussions, and Yong-il Shin for
experimental assistance. This work was supported by NSF and the
Office of Naval Research, AFOSR (through the MURI program), and
under Army Research Office grant no. W911NF-07-1-0493 with funds
from the DARPA Optical Lattice Emulator program.


\begin{thebibliography}{10}

\bibitem{MITRF}
S.~Gupta et~al.,
\newblock Science {\bf 300}, 1723 (2003).

\bibitem{JILARF}
C.~A.~Regal and D.~S.~Jin,
\newblock Phys. Rev. Lett. {\bf 90}, 230404 (2003).

\bibitem{MITBragg}
J.~Stenger et~al.,
\newblock Phys. Rev. Lett. {\bf 82}, 4569 (1999).

\bibitem{DemlerInterference}
T.~Kitagawa, A.~Aspect, M.~Greiner, and E.~Demler,
\newblock arXiv:1001.4358  (2010)

\bibitem{ZoranInterference}
Z.~Hadzibabic, S.~Stock, B.~Battelier, V.~Bretin, and J.~Dalibard,
\newblock Phys. Rev. Lett. {\bf 93}, 180403 (2004).

\bibitem{AspectHBT}
T.~Jeltes et~al.,
\newblock Nature {\bf 445}, 402 (2007).

\bibitem{Bloch06NoiseFermion}
T.~Rom et~al.,
\newblock Nature {\bf 444}, 733 (2006).

\bibitem{Jin05PairCorrelation}
M.~Greiner, C.~A. Regal, J.~T. Stewart, and D.~S. Jin,
\newblock Phys. Rev. Lett. {\bf 94}, 110401 (2005).

\bibitem{Chin09Incompressibility}
N.~Gemelke, X.~Zhang, C.~L. Hung, and C.~Chin,
\newblock Nature {\bf 460}, 995 (2009).

\bibitem{GreinerMottSingle}
W.~S. Bakr et~al.,
\newblock Science {\bf 329}, 547 (2010).

\bibitem{BlochMottSingle}
J.~F. Sherson et~al.,
\newblock Nature {\bf 467}, 68 (2010).

\bibitem{EsslingerFermiNoise}
T.~M\"uller et~al.,
\newblock Phys. Rev. Lett. {\bf 105}, 040401 (2010).

\bibitem{KetterleFermiNoise}
C.~Sanner et~al.,
\newblock Phys. Rev. Lett. {\bf 105}, 040402 (2010).

\bibitem{DuanSpinExchange}
L.-M. Duan, E.~Demler, and M.~D. Lukin,
\newblock Phys. Rev. Lett. {\bf 91}, 090402 (2003).

\bibitem{KuklovLattice}
A.~B. Kuklov and B.~V. Svistunov,
\newblock Phys. Rev. Lett. {\bf 90}, 100401 (2003).

\bibitem{StringariSpinFluc}
A.~Recati and S.~Stringari,
\newblock arXiv:1007.4504 (2010)

\bibitem{NoiseSpectroscopyDemler}
G.~M. Bruun, B.~M. Andersen, E.~Demler, and A.~S. S\o{}rensen,
\newblock Phys. Rev. Lett. {\bf 102}, 030401 (2009).

\bibitem{Ketterle98Varenna}
W.~Ketterle, D.~S. Durfee and D.~M. Stamper-Kurn,
\newblock Making, probing and understanding Bose-Einstein condensates,
\newblock in {\em Proceedings of the International School of Physics Enrico Fermi}, Varenna, 1998, (IOS, Amsterdam, 1999).

\bibitem{Goodman07}
J.~W. Goodman,
\newblock {\em Speckle Phenomena in Optics},
\newblock Ben Roberts and Company, Greenwood Village, CO, 2007.

\bibitem{ErtmerPRL}
D.~Hellweg et~al.,
\newblock Phys. Rev. Lett. {\bf 91}, 010406 (2003).

\bibitem{ScholtenDefocusContrast}
L.~D. Turner, K.~P. Weber, D.~Paganin, and R.~E. Scholten,
\newblock Opt. Lett. {\bf 29}, 232 (2004).

\bibitem{LevinCondensateFraction}
Q.~Chen, J.~Stajic, and K.~Levin,
\newblock Phys. Rev. Lett. {\bf 95}, 260405 (2005).

\bibitem{JILASweep}
M.~Greiner, C.~A.~Regal, and D.~S.~Jin,
\newblock Nature {\bf 426}, 537 (2003).

\bibitem{MITSweep}
M.~W. Zwierlein et~al.,
\newblock Phys. Rev. Lett. {\bf 91}, 250401 (2003).

\bibitem{AstrCompr}
G.~E. Astrakharchik, J.~Boronat, J.~Casulleras, and S.~Giorgini,
\newblock arXiv:cond-mat/0406113v1 (2004)

\bibitem{LandauStatMech}
E.~Lifshitz and L.~Pitaevskii,
\newblock {\em Statistical Physics Part 2},
\newblock (Pergamon Press Inc., NY, 1980),
\newblock L.D. Landau and E.M. Lifshitz, Course of Theoretical Physics, Vol. 9.

\bibitem{AstrQuantumNoise}
G.~E. Astrakharchik, R.~Combescot, and L.~P. Pitaevskii
\newblock Phys. Rev. A {\bf 76}, 063616 (2007).

\bibitem{HannafordStructureFactor}
E.~D. Kuhnle et~al.,
\newblock Phys. Rev. Lett. {\bf 105}, 070402 (2010).

\bibitem{SalomonEquationOfState}
N.~Navon, S.~Nascimbene, F.~Chevy, and C.~Salomon,
\newblock Science {\bf 328}, 729 (2010).

\bibitem{DeMarcoReview}
D.~McKay and B.~DeMarco,
\newblock arXiv:1010.0198 (2010)




\end{thebibliography}
\end{document}